\documentclass[]{spie}  %>>> use for US letter paper
\usepackage{amsmath,amsfonts,amssymb}
\usepackage{graphicx}
\usepackage[colorlinks=true, allcolors=blue]{hyperref}

\title{CCAT: Prime-Cam Optics Overview and Status Update}

\author[a]{Zachary B. Huber}
\author[a]{Lawrence T. Lin}
\author[a,b]{Eve M. Vavagiakis}
\author[c]{Rodrigo G. Freundt}
\author[d]{Victoria Butler}
\author[e,f]{Scott C. Chapman}
\author[g]{Steve K. Choi}
\author[a,c]{Abigail T. Crites}
\author[a]{Cody J. Duell}
\author[h]{Patricio A. Gallardo}
\author[i]{Anthony I. Huber}
\author[a]{Ben Keller}
\author[a]{Alicia Middleton}
\author[a,c]{Michael D. Niemack}
\author[j]{Thomas Nikola}
\author[k]{John Orlowski-Scherer}
\author[a]{Ema Smith}
\author[c]{Gordon Stacey}
\author[a,l]{Samantha Walker}
\author[m]{Bugao Zou}
\affil[a]{Department of Physics, Cornell University, Ithaca, NY, USA}
\affil[b]{Department of Physics, Duke University, Durham, NC, USA}
\affil[c]{Department of Astronomy, Cornell University, Ithaca, NY, USA}
\affil[d]{Cornell Laboratory for Accelerator-based Sciences and Education, Ithaca, NY, USA}
\affil[e]{Department of Physics and Astronomy, University of British Columbia, Vancouver, BC, Canada}
\affil[f]{Department of Physics and Atmospheric Science, Dalhousie University, Halifax, NS, Canada}
\affil[g]{Department of Physics and Astronomy, University of California, Riverside, CA, USA}
\affil[h]{Kavli Institute for Cosmological Physics, University of Chicago, Chicago, IL, USA}
\affil[i]{Department of Physics and Astronomy, University of Victoria, Victoria, BC, Canada}
\affil[j]{Cornell Center for Astrophysics and Planetary Sciences, Ithaca, NY, USA}
\affil[k]{Department of Physics and Astronomy, University of Pennsylvania, Philadelphia, PA, USA}
\affil[l]{Cornell Center for Materials Research, Ithaca, NY, USA}
\affil[m]{Department of Applied and Engineering Physics, Cornell University, Ithaca, NY, USA}

\authorinfo{Further author information: (Send correspondence to Z.B.H.)\\Z.B.H.: E-mail: zbh5@cornell.edu}

\begin{document} 
\maketitle

\begin{abstract}
Prime-Cam is a first-generation science instrument for the CCAT Observatory's six-meter aperture Fred Young Submillimeter Telescope (FYST). FYST’s crossed-Dragone design provides high optical throughput to take advantage of its unique site at 5600 m on Cerro Chajnantor in Chile’s Atacama Desert to reach mapping speeds over ten times greater than current and near-term submillimeter experiments. Housing up to seven independent instrument modules in its 1.8-meter diameter cryostat, Prime-Cam will combine broadband polarization-sensitive modules and spectrometer modules designed for observations in several frequency windows between 210 GHz and 850 GHz to study a wide range of astrophysical questions from Big Bang cosmology to the formation of stars and galaxies in the Epoch of Reionization and beyond. In order to cover this range of frequencies and observation modes, each of the modules contains a set of cold reimaging optics that is optimized for the science goals of that module. These optical setups include several filters, three or four anti-reflection-coated silicon lenses, and a Lyot stop to control the field of view and illumination of the primary mirror, satisfy a series of mechanical constraints, and maximize optical performance within each passband. We summarize the design considerations and trade-offs for the optics in these modules and provide a status update on the fabrication of the Prime-Cam receiver and the design of its 1 K and 100 mK thermal BUSs. 
\end{abstract}


\section{INTRODUCTION}
\label{sec:intro} 

The CCAT Collaboration is constructing the Fred Young Submillimeter Telescope (FYST) \cite{Parshley2018} at 5600 m on Cerro Chajnantor in the Atacama Desert in Chile to probe cosmology and the history of galaxy evolution and galactic ecosystems across a wide range of cosmic time. In order to study early dusty star-forming galaxies, the Epoch of Reionization from redshifts of 4 to 8, galaxy clusters via the Sunyaev-Zel'dovich effect, thermal dust emission in the Milky Way and other galaxies, and much more across several millimeter (mm) and sub-mm bands \cite{ccat_science_2021}, FYST requires an excellent instrument that can take advantage of the large field of view and throughput of its crossed-Dragone design with a six-meter primary mirror \cite{Niemack_telescope_2016}. Prime-Cam is one of the first-generation science instruments on FYST, and it will be able to hold seven independent instrument modules in its 1.8-m diameter cryostat when fully populated \cite{Vavagiakis_PrimeCam_2018} \cite{Choi_CCATsensitivity_2020}. Initially, Prime-Cam will be deployed with three broadband, polarization-sensitive modules with passbands centered around 280 GHz \cite{VavagiakisModCam2022}, 350 GHz, and 850 GHz \cite{Chapman_850GHz_2022} and one spectrometer module, the Epoch of Reionization Spectrometer (EoR-Spec) \cite{Nikola_EoRSpecUpdate_2022}.

Section~\ref{sec:fab} provides an update on the fabrication status of the Prime-Cam cryostat. The current state of the design for the 1 K and 100 mK thermal BUSs is summarized in section~\ref{sec:cryo}, which also outlines a possible method for inserting and removing modules horizontally. Section~\ref{sec:optics} summarizes the status of the designs for the cold optics of the four initial instrument modules with a particular focus on the design process and outcomes for the 350 GHz and EoR-Spec modules, and section~\ref{sec:summary} describes our next steps.

\section{PRIME-CAM FABRICATION UPDATE}
\label{sec:fab}
\begin{figure}[b]
    \centering
    \includegraphics[scale=0.5]{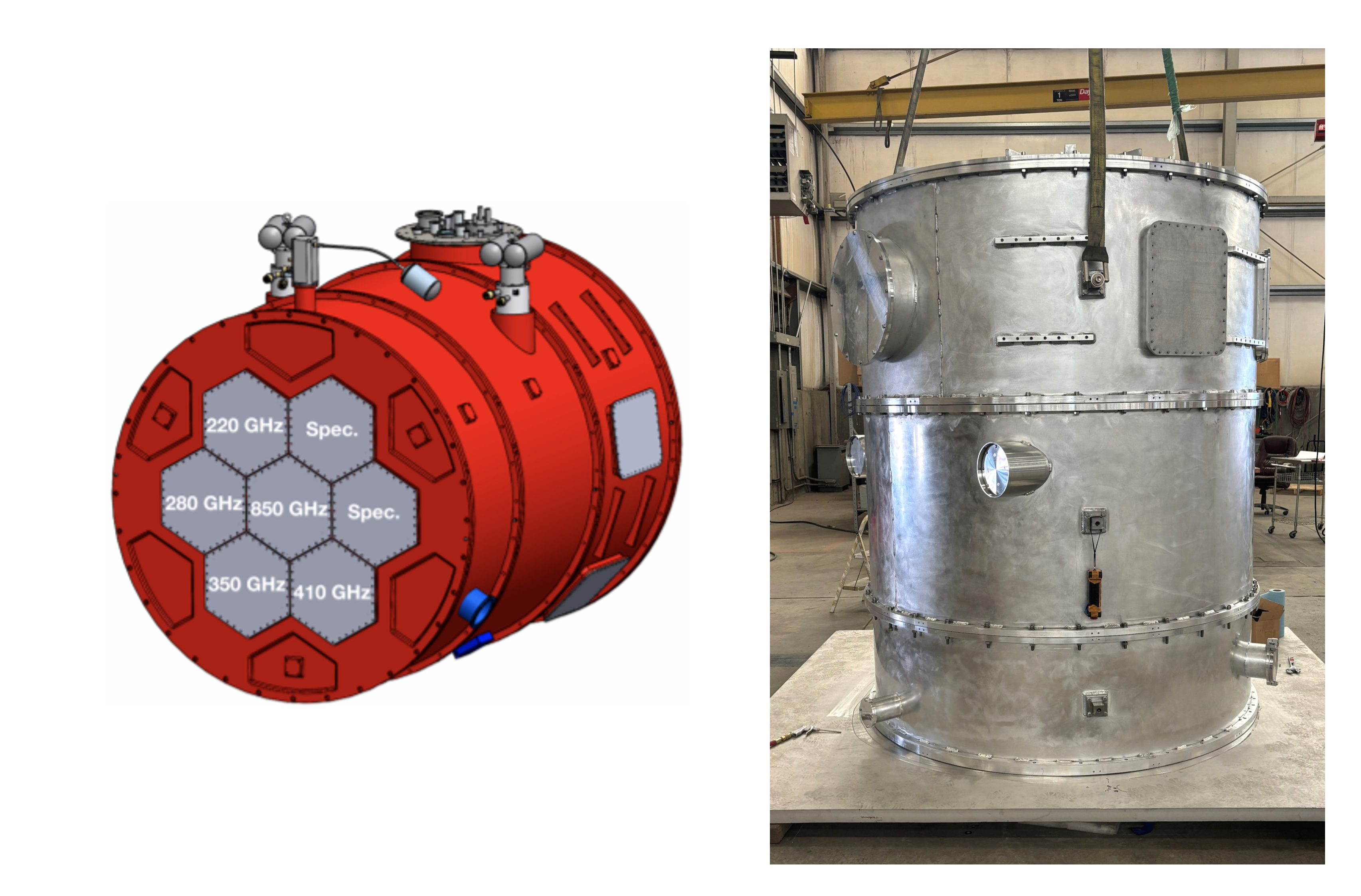}
    \caption{\textit{Left}: The design of the Prime-Cam cryostat showing planned locations of the different modules from Ref.~\citenum{Choi_CCATsensitivity_2020}. \textit{Right}: The Prime-Cam cryostat in a mostly assembled state in the Redline Chambers facility in Utah. The ports for the DR and PTs can be seen on the left side, while a readout harness port can be seen to the top right.}
    \label{fig:primecam_fab}
\end{figure}
Prime-Cam is currently being fabricated by Redline Chambers in Utah. The major shells have been machined and fit checked (see Fig.~\ref{fig:primecam_fab}), and final machining and welding of the front window plate and minor shells is ongoing. Prime-Cam's internal epoxied G10 support tabs are finished and ready to be installed. Once the vacuum shell is complete, fit checks and vacuum leak testing will be performed before the cryostat is delivered to Cornell for testing. Interior components such as the thermal connections to the pulse tube (PT) 80 K, 40 K and 4 K cold heads and dilution refrigerator (DR) cold stages, including custom adapters and copper straps from Technology Applications, Inc.~(TAI), are in fabrication. We expect Prime-Cam to be delivered by the end of 2024.

Mod-Cam, a single instrument module testbed for Prime-Cam, is in testing at Cornell \cite{VavagiakisModCam2022}. Mod-Cam enables faster swapping of instrument modules than Prime-Cam, and all Prime-Cam instrument modules are designed to be compatible with Mod-Cam. Mod-Cam is currently dark testing the 280 GHz module, and later this year, optical testing of the module will begin.

\section{CRYOGENIC THERMAL BUS DESIGN}
\label{sec:cryo}

\begin{figure}
    \centering
    \includegraphics[scale=0.55]{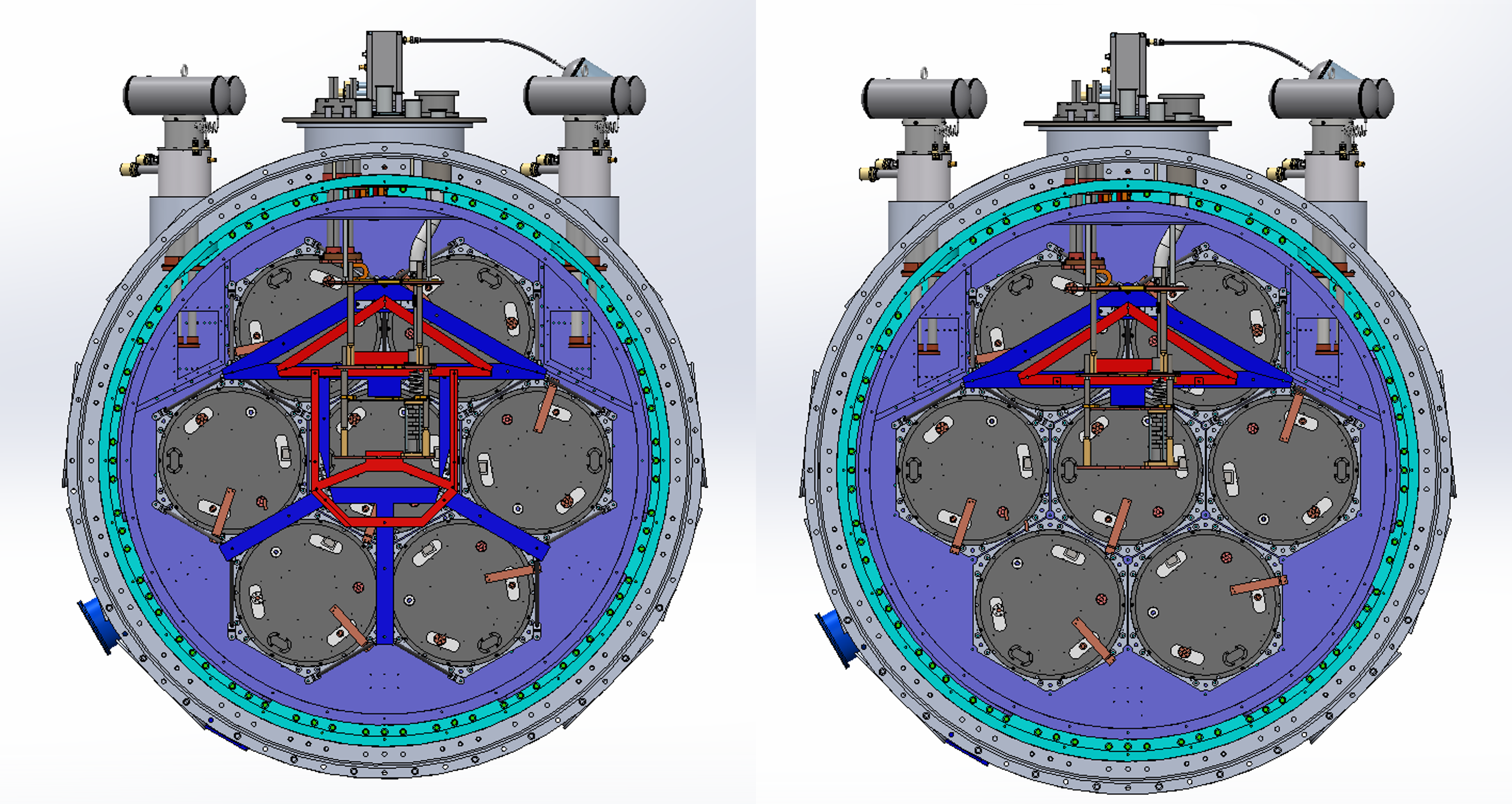}
    \caption{\textit{Left}: The current design of the thermal BUS for Prime-Cam showing both the 1 K BUS (blue) and 100 mK BUS (red). The rectangular extrusions from the main frame of each thermal BUS are the locations for connecting the straps to the BUS. The orientation of the module 1 K and 100 mK cold fingers is set by the Mod-Cam test cryostat design. \textit{Right}: The thermal BUS with the bottom part removed so that the bottom four instrument modules can be installed and removed without removing the Prime-Cam DR. The DR (center of the picture) would still block the central module.}
    \label{fig:thermal_bus}
\end{figure}

While Prime-Cam's fabrication proceeds, design work continues in parallel for internal cryogenic components like the thermal BUSs and copper straps. Prime-Cam relies on a single Bluefors DR to provide cooling power at 1 K and 100 mK and three supplemental Cryomech PTs, one PT-90 to cool the 80 K stage and two PT-420s to cool the 40 K and 4 K stages. The 4 K stage of each instrument module will be cooled through its interface with the 4 K stage of Prime-Cam, which is also how the modules are mounted into Prime-Cam. The 1 K and 100 mK stages, meanwhile, need to be cooled by coupling to the relevant stages of the DR. The 1 K thermal BUS and 100 mK thermal BUS are large oxygen-free high-conductivity (OFHC) copper structures that are mounted by thermally-isolating carbon fiber trusses at the back of Prime-Cam (see left of Fig.~\ref{fig:thermal_bus}). Copper straps will connect these thermal BUSs to the DR stages and to the cold fingers that protrude from the rear of each instrument module to distribute cooling power from the DR to the relevant module temperature stages.

Because the thermal BUSs and DR sit behind the modules, they would need to be removed in order to extract or insert a module from the back of Prime-Cam. To minimize the need for such a complex and time-consuming maneuver, we are exploring designs in which the thermal BUSs are separated into two parts, a top and a bottom, such that the lower part can be removed without taking the DR out of Prime-Cam. Such designs could allow for the bottom four instrument modules in Prime-Cam to be removed or inserted much more easily and quickly. The right side of Fig.~\ref{fig:thermal_bus} shows what the current thermal BUS design would look like with the bottom removed. In order to mitigate thermal conductivity issues around the joints that connect the two parts of the thermal BUS, we plan to run separate copper straps from the DR to both the top and the bottom of each thermal BUS. Thermal and mechanical finite element analyses (FEAs) are underway to assess the efficacy and stability of this design and possible design variants.

\begin{figure}
    \centering
    \includegraphics[width=0.9\linewidth]{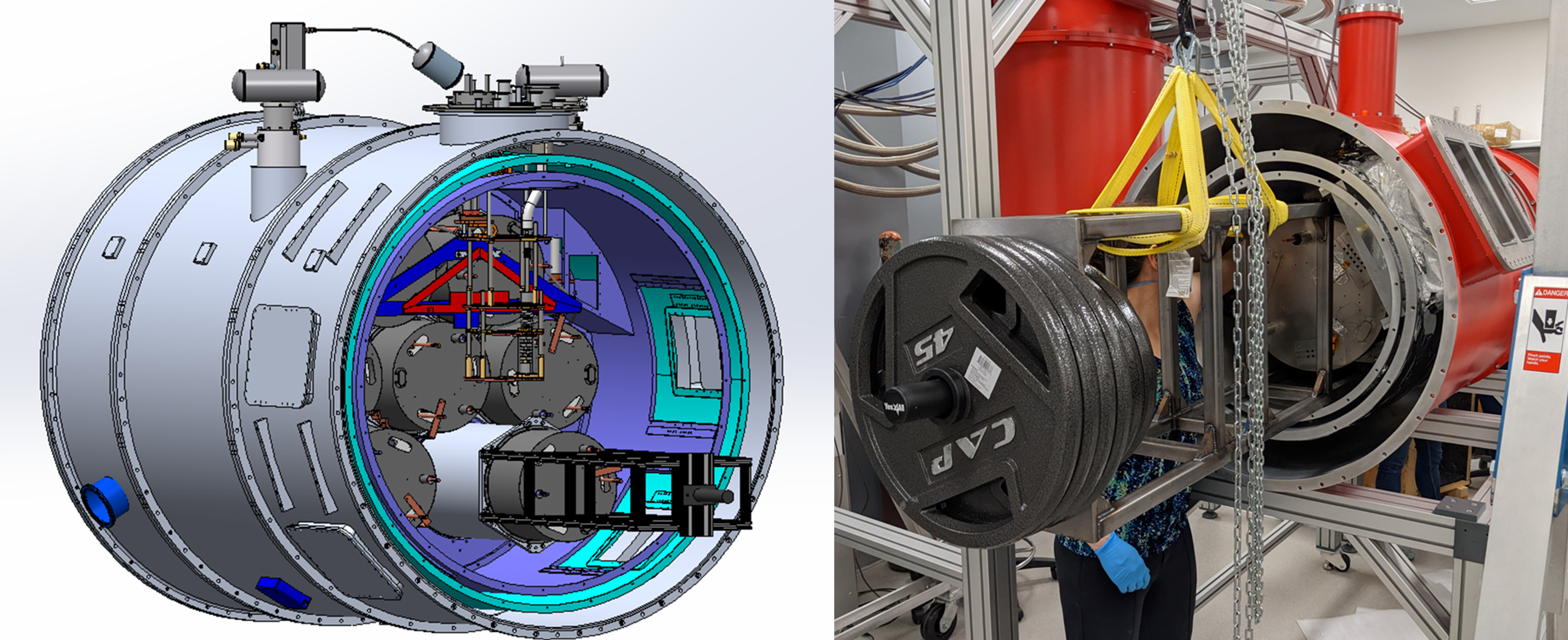}
    \caption{\textit{Left}: A CAD rendering showing Prime-Cam with the battering ram attached to one of the lower modules. With the bottom half of the thermal BUSs removed, the lower four modules may be removed via this method. \textit{Right}: A photograph of the battering ram being used to install the 280 GHz module into Mod-Cam.}
    \label{fig:battering_ram}
\end{figure}

In order to take advantage of this modular design, we need a way to insert and remove modules horizontally into Prime-Cam. To accomplish this, we have designed a steel ``battering ram" -- a steel frame that mounts to the 4 K flange of the instrument module. The left side of Fig.~\ref{fig:battering_ram} shows a CAD rendering of the current design and what this might look like when removing one of the lower modules in Prime-Cam. Extra weight can be added to the back of the battering ram to counter-balance the weight of the module, and the entire battering ram can be lifted with a gantry crane. We have built and tested this design for our single-module test cryostat, Mod-Cam \cite{VavagiakisModCam2022}, as shown on the right side of Fig.~\ref{fig:battering_ram}. We have had success using this method in Mod-Cam, but testing will need to be done to assess the feasibility of this method for the larger Prime-Cam cryostat.

\section{INSTRUMENT MODULE COLD OPTICS}
\label{sec:optics}
The designs of the cryogenic reimaging optics for the four funded instrument modules that will initially populate Prime-Cam vary based on the different science requirements of the modules. For ease of fabrication, the 280 GHz module uses the same optical design as the instrument modules for the Simons Observatory (SO).\cite{VavagiakisModCam2022}\textsuperscript{,}\cite{Dicker2018}\textsuperscript{,}\cite{Gallardo_SO_optics_2018} Using the same optics as SO also will allow us to compare our optical performance and any systematic effects more readily and to contribute to some of the same science goals as the 220/280 GHz modules for the SO Large Aperture Telescope Receiver (LATR). 

The 850 GHz module has the most stringent optical requirements of the planned Prime-Cam instrument modules due to the smaller beam size and mirror surface precision requirements at this higher frequency compared to our other instrument modules. For this reason, it will be located in the central position in Prime-Cam, which has the highest optical performance in the telescope focal plane. In order to maintain excellent image quality in this frequency band while maximizing the field of view, the 850 GHz module's optics design incorporates a fourth lens in a unique configuration as described in Ref.~\citenum{AHuber2022}.

The 350 GHz module is similar to the 280 GHz module and shares many of its science goals and technical requirements, though the higher frequency band opens up windows into other science cases, including more thorough studies of thermal dust emission. To expedite the construction of this module, we are using the same mechanical design as the 280 GHz module. If we were to use the optical design of the 280 GHz module without reoptimizing for this higher frequency band, it would result in a degradation of optical quality. The 280 GHz design covers greater than 90\% of the detectors in the focal plane with a Strehl ratio above 0.8 in the 280 GHz band, which we consider diffraction-limited. In the 350 GHz band, the same design only covers 70-80\% of the detectors with a Strehl ratio above 0.8 for the locations in Prime-Cam in which we are considering placing the 350 GHz module, though more than 95\% of the detectors would see a Strehl ratio above 0.7. 

\begin{figure}
    \centering
    \includegraphics[width=0.9\linewidth]{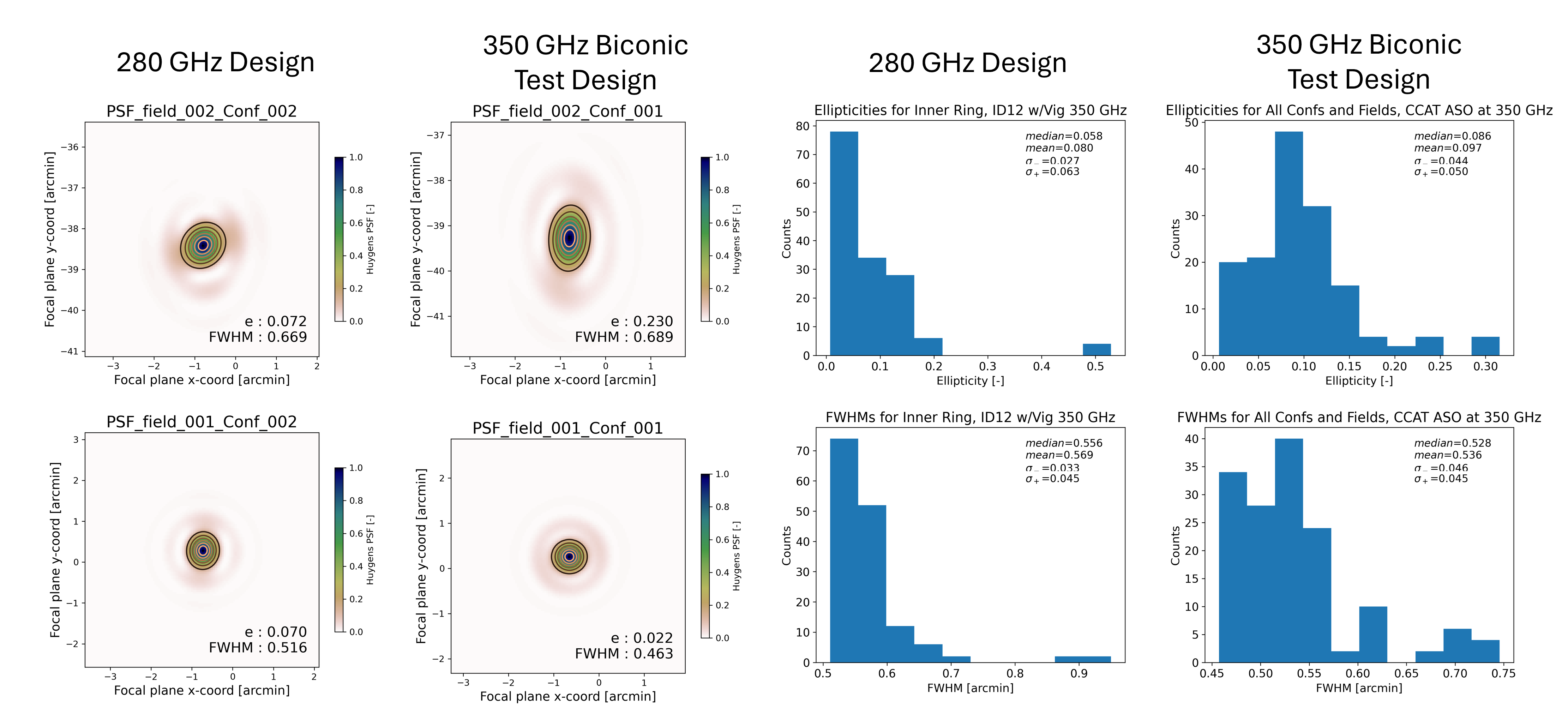}
    \caption{\textit{Left}: Example PSF plots fit with a 2D Gaussian to extract the FWHM and ellipticity (e) of the main beam. The left column shows the 280 GHz module optics design PSF at the edge of the focal plane (top) and the center (bottom), while the right column shows the same thing for the best of the alternative biconic designs that were considered for the 350 GHz module. Despite a different numbering convention, all plots are for the same receiver location. \textit{Right}: Histograms of the FWHM and ellipticity for both designs. The data come from sampling the PSF in twenty-five points across the focal plane for each of the six possible locations for the module. The median FWHM improves for the biconic design, but at the expense of an increased ellipticity across the focal plane.}
    \label{fig:350GHz_psfs}
\end{figure}
As such, we considered a series of alternative designs to change the shapes of the lenses without altering their mechanical positions. The best optical performance of any design that we considered resulted from changing one of the radially-symmetric aspheric lenses to a biconic lens, which has differently shaped cross-sections along two perpendicular axes. This design would have resulted in a roughly 10\% increase in the number of detectors that had a diffraction-limited Strehl ratio, but an analysis with the point spread function (PSF) tools in Zemax revealed that this improvement in Strehl ratio was accompanied by additional variance in the ellipticity of the PSF. The left side of Fig.~\ref{fig:350GHz_psfs} shows examples of a few of these PSF plots. While the PSF size and ellipticity improved for the biconic design in the center of the focal plane, the middle and edges of the focal plane became much more elliptical. The right side of Fig.~\ref{fig:350GHz_psfs} shows histograms of the full width at half maximum (FWHM) and ellipticity for each design. These histograms use data from all six outer positions in Prime-Cam with the PSFs sampled at twenty-five regularly spaced locations across the focal plane at each position. While the biconic design did reduce the worst outliers in both FWHM and ellipticity, it also caused the median ellipticity to increase while reducing the median FWHM compared to the all-aspheric design. This reduction in FWHM is what caused the Strehl ratio coverage to increase, though the extra ellipticity limited the optimization from achieving even higher Strehl ratios across the focal plane. These effects were true in the aggregate for all six possible module locations and for each possible module location individually.

The increase in the median ellipticity and the variation of the ellipticity across the focal plane would have made it much harder to correct for the effects of the beam in the maps made with this module. We decided to reduce the risk of additional systematic effects on our data and the risk of schedule delay from developing new metamaterial anti-reflection coatings \cite{Datta_ARcoatings_2013} for the biconic lenses by adopting the 280 GHz module optics design for the 350 GHz module as well. The lenses are now being fabricated for this module. Given that we were optimizing new designs for this module under tight constraints from using the same mechanical design as the 280 GHz module, it is quite possible that improvements could be made to the optical design for a future upgrade to this module if necessary. In the short term, however, this choice reduces the complexity of this module and mitigates a substantial amount of risk in the design, module assembly, and data analysis.

\begin{figure}
    \centering
    \includegraphics[width=1.0\linewidth]{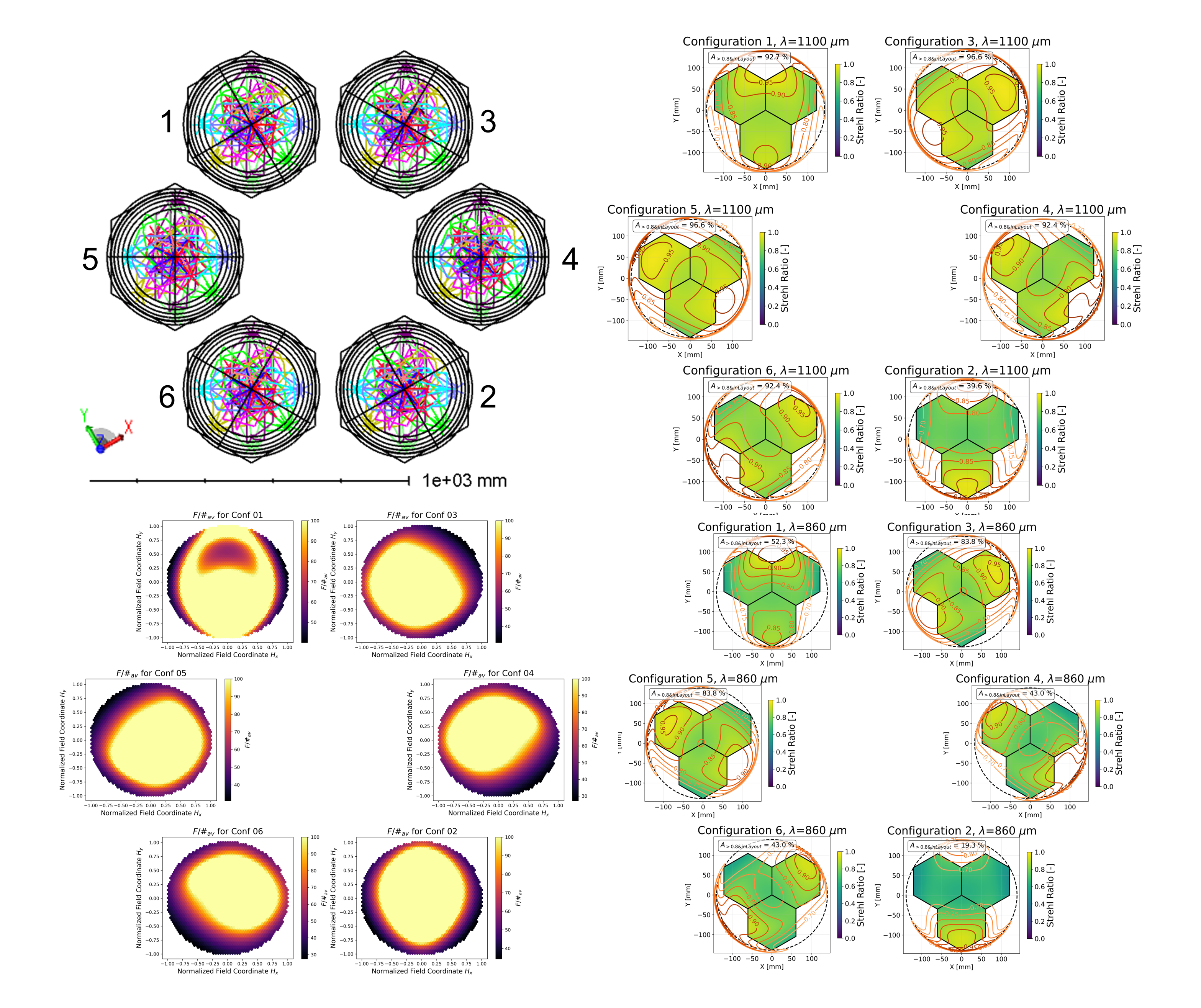}
    \caption{\textit{Top Left}: An image from the Zemax design of the six outer tubes in Prime-Cam (from Ref.~\citenum{huber_EoRSpec_optics_2022}). The colored lines show light rays being traced from different fields on the sky. The 850 GHz module will occupy the central position (not shown), while the other modules will go into the ring around the center. The tubes are labeled with their configuration number for identifying locations in the other plots in this figure. \textit{Top Right}: The Strehl ratio at 1.1 mm (270 GHz band) for the six outer tube locations in the new EoR-Spec design. The Zemax design is rotated 30 degrees relative to the normal Prime-Cam orientation to simulate observing at an elevation of 60 degrees, which is why each plot looks rotated relative to the symmetry axis of the telescope mirrors. Configuration 5 is the planned location for EoR-Spec, and over 96\% of the detectors have a Strehl ratio above 0.8 there. \textit{Bottom Left}: The F/\# at the Lyot stop for each possible EoR-Spec location in Prime-Cam. The colorbar cuts off at 100, the target for this metric of the design. The majority of the focal plane meets our collimation requirements for each configuration. \textit{Bottom Right}: The Strehl ratio at 0.86 mm (360 GHz band) for the six outer tube locations in the new EoR-Spec design. Configuration 5 is the planned location for EoR-Spec, and over 83\% of the detectors have a Strehl ratio above 0.8 there. Only one of the detector arrays will be sensitive to this band, so this location in Prime-Cam exceeds our requirements for optical performance.}
    \label{fig:eor-spec_optics}
\end{figure}
The design for EoR-Spec is necessarily more complicated than any of the broadband modules because of the Fabry-Perot interferometer (FPI) located at the Lyot stop of the system. To maximize the performance of the FPI, the beam must be highly collimated at the Lyot stop and must be refocused in a relatively short space onto the focal plane. Previous development work led to a design that made use of three aspheric lenses and one biconic lens \cite{huber_EoRSpec_optics_2022}. EoR-Spec will have two detector arrays with a passband center of 270 GHz, and one with a passband center of 360 GHz, and this design was able to achieve diffraction-limited performance for well above two-thirds of the focal-plane at 270 GHz and for about one-third of the focal plane at 360 GHz in most possible locations of the module in Prime-Cam. It also maintained good collimation at the Lyot stop for the majority of fields on the sky as measured by calculating the F/\# for a bundle of four extreme light rays traced from different locations on the sky.

The design work for the optics for the 350 GHz module and the modules of the Advanced Simons Observatory (ASO) upgrade to SO's LATR, however, seemed to suggest that it may not be necessary to use more expensive and complicated biconic lens for EoR-Spec. The biconic lenses are valuable to correct for the astigmatism of the telescope mirrors, which is the leading order optical aberration of the coma-corrected crossed-Dragone design. The astigmatism becomes more pronounced further away from the telescope's main optical axis, which is equivalent to further away from the center of the receiver. For the LATR, which is larger than Prime-Cam, it was clear that biconic lenses were a substantial improvement at the edges of the cryostat, but the gain was less substantial in the area covered by Prime-Cam's instrument modules. The 350 GHz module optical design work also seemed to suggest that using biconic lenses in Prime-Cam might not be as valuable as once thought, so we resumed exploring the parameter space of designs with four aspheric lenses.

\begin{figure}[b]
    \centering
    \includegraphics[width=1.0\linewidth]{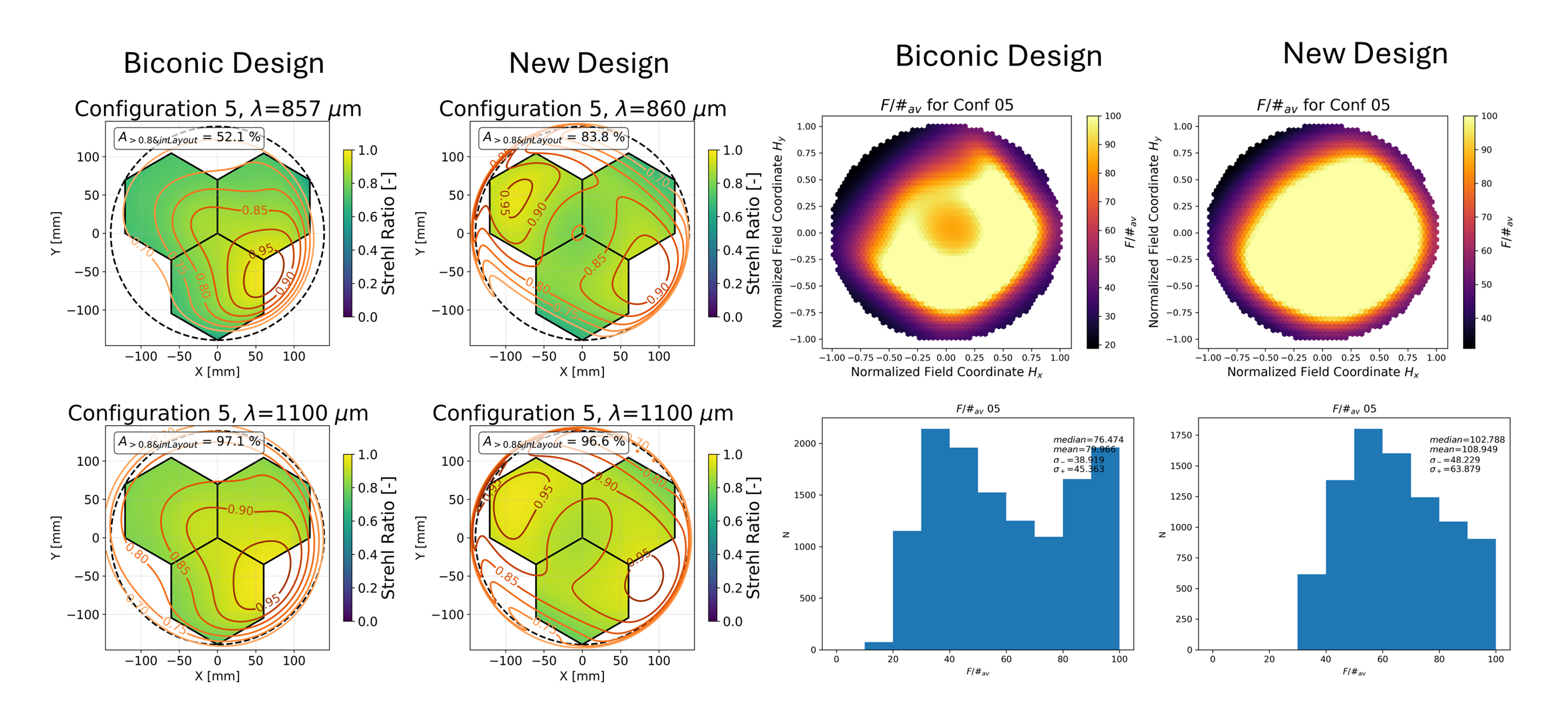}
    \caption{\textit{Left}: A comparison of the Strehl ratio at the likely location in Prime-Cam between the previous best biconic design (left column) and the new all-aspheric design (right column) at the 360 GHz band (0.86 mm; top row) and the 270 GHz band (1.1 mm; bottom row). The performance at 270 GHz is similiar between the two designs while the performance at higher frequencies is improved.  \textit{Right}: A comparison of the F/\# at the stop between the previous best biconic design (left column) and the new all-aspheric design (right column). The histograms on the bottom are limited at 100 to show the performance of the pixels that do not meet that target, but the median and mean reported in the figure is for the full data set. The increase in the median value from the biconic design to the all-aspheric design indicates that the collimation at the FPI is improved in the all-aspheric design.}
    \label{fig:eor-spec_optics_comparison}
\end{figure}
By paying more careful attention to controlling the illumination of the primary mirror and using more accurate ray tracing tools within Zemax throughout the optimization process, we were able to develop an all-aspheric design for EoR-Spec that had similar optical performance in the 270 GHz band, improved optical performance in the 360 GHz band, and slightly improved performance in the collimation at the stop. Fig.~\ref{fig:eor-spec_optics} shows the optical performance of the new design for all three of these metrics across all possible locations in Prime-Cam, while Fig.~\ref{fig:eor-spec_optics_comparison} shows a direct comparison of the biconic design presented in Ref.~\citenum{huber_EoRSpec_optics_2022} and the all-aspheric design that we adopted for Configuration 5, the planned location of EoR-Spec. This improved and simplified design was recently finalized and is now being fabricated.

\section{SUMMARY}
\label{sec:summary}

The Prime-Cam cryostat and the first four instrument modules that will populate it are in various stages of fabrication with Prime-Cam expected to arrive at Cornell later this year. Preliminary plans are in place for the thermal BUS at 1 K and 100 mK to distribute cooling power to Prime-Cam's modules while utilizing a modular design to allow the bottom four modules to be removed without removing the DR. Thermal and mechanical analyses of this design will lead to further refinements to the BUSs and other cryogenic components as we prepare for fabrication and cryogenic testing in Prime-Cam.

The optical designs for EoR-Spec and the 350 GHz module were finalized recently and submitted to a vendor for machining of the silicon lenses. The 350 GHz module adopts the same optical design as the 280 GHz module for ease and speed of fabrication; alternative designs that maintained the same lens positions to use the same mechanical design as the 280 GHz module did not provide substantial improvement due to larger PSF ellipticity while also posing greater cost and effort burdens. The EoR-Spec optical design was improved using lessons learned from the 350 GHz module optical design process, permitting the use of four aspheric lenses instead of a design that used a biconic lens while slightly improving performance. These designs have been submitted for machining along with the mechanical components. Progress is also being made on the other optical elements, detector wafers, and readout equipment for EoR-Spec \cite{Freundt_EoRSpec_2024}, the 850 GHz module \cite{AHuber_850GHz_2024}, and the 350 GHz module. The 280 GHz module is largely complete and undergoing cryogenic testing in Mod-Cam at Cornell. We plan to assemble and test all four of the initial modules for Prime-Cam in the coming year.

\acknowledgments % equivalent to \section*{ACKNOWLEDGMENTS}       

ZBH would like to thank PAG for software for making the Zemax plots and PAG and Simon Dicker for advice on designing and optimizing optical systems.
 
The CCAT project, FYST and Prime-Cam instrument have been supported by generous contributions from the Fred M. Young, Jr. Charitable Trust, Cornell University, and the Canada Foundation for Innovation and the Provinces of Ontario, Alberta, and British Columbia. The construction of the FYST telescope was supported by the Gro{\ss}ger{\"a}te-Programm of the German Science Foundation (Deutsche Forschungsgemeinschaft, DFG) under grant INST 216/733-1 FUGG, as well as funding from Universit{\"a}t zu K{\"o}ln, Universit{\"a}t Bonn and the Max Planck Institut f{\"u}r Astrophysik, Garching. The construction of EoR-Spec is supported by NSF grant AST-2009767. The construction of the 350 GHz instrument module for Prime-Cam is supported by NSF grant AST-2117631. 

ZBH acknowledges funding from a NASA Space Technology Graduate Research Opportunities award. EMV acknowledges support from NSF award AST-2202237.

% References
\bibliography{report} % bibliography data in report.bib

\begin{thebibliography}{10}

\bibitem{Parshley2018}
Parshley, S.~C., Niemack, M., Hills, R., Dicker, S.~R., D{\"u}nner, R., Erler, J., Gallardo, P.~A., Gudmundsson, J.~E., Herter, T., Koopman, B.~J., Limon, M., Matsuda, F.~T., Mauskopf, P., Riechers, D.~A., Stacey, G.~J., and Vavagiakis, E.~M., ``Cold optical design for the {L}arge {A}perture {S}imons {O}bservatory telescope,'' in [{\em Ground-based and Airborne Telescopes VII}{\nolinebreak\hspace{0.1em}]},  {\em Proc. SPIE} {\bf 10700},  1292--1304 (2018).

\bibitem{ccat_science_2021}
{CCAT-Prime Collaboration}, Aravena, M., Austermann, J.~E., Basu, K., Battaglia, N., Beringue, B., Bertoldi, F., Bigiel, F., Bond, J.~R., Breysse, P.~C., Broughton, C., Bustos, R., Chapman, S.~C., Charmetant, M., Choi, S.~K., Chung, D.~T., Clark, S.~E., Cothard, N.~F., Crites, A.~T., Dev, A., Douglas, K., Duell, C.~J., Ebina, H., Erler, J., Fich, M., Fissel, L.~M., Foreman, S., Gao, J., García, P., Giovanelli, R., Haynes, M.~P., Hensley, B., Herter, T., Higgins, R., Huber, Z., Hubmayr, J., Johnstone, D., Karoumpis, C., Keating, L.~C., Komatsu, E., Li, Y., Magnelli, B., Matthews, B.~C., Meerburg, P.~D., Meyers, J., Muralidhara, V., Murray, N.~W., Niemack, M.~D., Nikola, T., Okada, Y., Riechers, D.~A., Rosolowsky, E., Roy, A., Sadavoy, S.~I., Schaaf, R., Schilke, P., Scott, D., Simon, R., Sinclair, A.~K., Sivakoff, G.~R., Stacey, G.~J., Stutz, A.~M., Stutzki, J., Tahani, M., Thanjavur, K., Timmermann, R.~A., Ullom, J.~N., van Engelen, A., Vavagiakis, E.~M., Vissers, M.~R., Wheeler, J.~D., White, S. D.~M., Zhu,
  Y., and Zou, B., ``Ccat-prime collaboration: Science goals and forecasts with prime-cam on the fred young submillimeter telescope,'' (2021).

\bibitem{Niemack_telescope_2016}
Niemack, M.~D., ``Designs for a large-aperture telescope to map the cmb 10x faster,'' {\em Appl. Opt.}~{\bf 55},  1688--1696 (Mar 2016).

\bibitem{Vavagiakis_PrimeCam_2018}
Vavagiakis, E.~M., Ahmed, Z., Ali, A., Basu, K., Battaglia, N., Bertoldi, F., Bond, R., Bustos, R., Chapman, S.~C., Chung, D., Coppi, G., Cothard, N.~F., Dicker, S., Duell, C.~J., Duff, S.~M., Erler, J., Fich, M., Galitzki, N., Gallardo, P.~A., Henderson, S.~W., Herter, T.~L., Hilton, G., Hubmayr, J., Irwin, K.~D., Koopman, B.~J., McMahon, J., Murray, N., Niemack, M.~D., Nikola, T., Nolta, M., Orlowski-Scherer, J., Parshley, S.~C., Riechers, D.~A., Rossi, K., Scott, D., Sierra, C., Silva-Feaver, M., Simon, S.~M., Stacey, G.~J., Stevens, J.~R., Ullom, J.~N., Vissers, M.~R., Walker, S., Wollack, E.~J., Xu, Z., and Zhu, N., ``{Prime-Cam: a first-light instrument for the CCAT-prime telescope},'' in [{\em Millimeter, Submillimeter, and Far-Infrared Detectors and Instrumentation for Astronomy IX}{\nolinebreak\hspace{0.1em}]},  Zmuidzinas, J. and Gao, J.-R., eds.,  {\bf 10708},  107081U, International Society for Optics and Photonics, SPIE (2018).

\bibitem{Choi_CCATsensitivity_2020}
Choi, S.~K., Austermann, J., Basu, K., Battaglia, N., Bertoldi, F., Chung, D.~T., Cothard, N.~F., Duff, S., Duell, C.~J., Gallardo, P.~A., Gao, J., Herter, T., Hubmayr, J., Niemack, M.~D., Nikola, T., Riechers, D., Rossi, K., Stacey, G.~J., Stevens, J.~R., Vavagiakis, E.~M., Vissers, M., and Walker, S., ``Sensitivity of the prime-cam instrument on the ccat-prime telescope,'' {\em Journal of Low Temperature Physics}~{\bf 199},  1089–1097 (Mar 2020).

\bibitem{VavagiakisModCam2022}
Vavagiakis, E.~M., Duell, C.~J., Austermann, J., Beall, J., Bhandarkar, T., Chapman, S.~C., Choi, S.~K., Coppi, G., Dicker, S., Devlin, M., Freundt, R.~G., Gao, J., Groppi, C., Herter, T.~L., Huber, Z.~B., Hubmayr, J., Johnstone, D., Keller, B., Kofman, A.~M., Li, Y., Mauskopf, P., McMahon, J., Moore, J., Murphy, C.~C., Niemack, M.~D., Nikola, T., Orlowski-Scherer, J., Rossi, K.~M., Sinclair, A.~K., Stacey, G.~J., Ullom, J., Vissers, M., Wheeler, J., Xu, Z., Zhu, N., and Zou, B., ``{CCAT-prime: design of the Mod-Cam receiver and 280 GHz MKID instrument module},'' in [{\em Millimeter, Submillimeter, and Far-Infrared Detectors and Instrumentation for Astronomy XI}{\nolinebreak\hspace{0.1em}]},  Zmuidzinas, J. and Gao, J.-R., eds.,  {\bf 12190},  1219004, International Society for Optics and Photonics, SPIE (2022).

\bibitem{Chapman_850GHz_2022}
Chapman, S.~C., Huber, A.~I., Sinclair, A.~K., Wheeler, J.~D., Austermann, J.~E., Beall, J., Burgoyne, J., Choi, S.~K., Crites, A., Duell, C.~J., Devina, J., Gao, J., Fich, M., Henke, D., Herter, T., Johnstone, D., Knee, L. B.~G., Niemack, M.~D., Rossi, K.~M., Stacey, G., Tsuchitori, J., Ullom, J., Lanen, J.~V., Vavagiakis, E.~M., and Vissers, M., ``{CCAT-prime: the 850 GHz camera for prime-cam on FYST},'' in [{\em Millimeter, Submillimeter, and Far-Infrared Detectors and Instrumentation for Astronomy XI}{\nolinebreak\hspace{0.1em}]},  Zmuidzinas, J. and Gao, J.-R., eds.,  {\bf 12190},  1219005, International Society for Optics and Photonics, SPIE (2022).

\bibitem{Nikola_EoRSpecUpdate_2022}
Nikola, T., Choi, S.~K., Duell, C.~J., Freundt, R.~G., Huber, Z.~B., Li, Y., Malavalli, K., Niemack, M., Rossi, K.~M., Stacey, G.~J., Vavagiakis, E.~M., Zou, B., Cothard, N.~F., Austermann, J., Wheeler, J.~D., Gao, J., Vissers, M.~R., Hubmayr, J., Beall, J., and Ullom, J., ``{CCAT-prime: the epoch reionization spectrometer for primce-cam on FYST},'' in [{\em Millimeter, Submillimeter, and Far-Infrared Detectors and Instrumentation for Astronomy XI}{\nolinebreak\hspace{0.1em}]},  Zmuidzinas, J. and Gao, J.-R., eds.,  {\bf 12190},  121900G, International Society for Optics and Photonics, SPIE (2022).

\bibitem{Dicker2018}
Dicker, S.~R., Gallardo, P.~A., Gudmundsson, J.~E., Mauskopf, P.~D., Ali, A., Ashton, P.~C., Coppi, G., Devlin, M.~J., Galitzki, N., Ho, S.~P., Hill, C.~A., Hubmayr, J., Keating, B., Lee, A.~T., Limon, M., Matsuda, F., McMahon, J., Niemack, M.~D., Orlowski-Scherer, J.~L., Piccirillo, L., Salatino, M., Simon, S.~M., Staggs, S.~T., Thornton, R., Ullom, J.~N., Vavagiakis, E.~M., Wollack, E.~J., Xu, Z., and Zhu, N., ``Cold optical design for the {L}arge {A}perture {S}imons {O}bservatory telescope,'' in [{\em Ground-based and Airborne Telescopes VII}{\nolinebreak\hspace{0.1em}]},  {\em Proc. SPIE} {\bf 10700},  1064--1076 (2018).

\bibitem{Gallardo_SO_optics_2018}
Gallardo, P.~A., Gudmundsson, J., Koopman, B.~J., Matsuda, F.~T., Simon, S.~M., Ali, A., Bryan, S., Chinone, Y., Coppi, G., Cothard, N., Devlin, M.~J., Dicker, S., Fabbian, G., Galitzki, N., Hill, C.~A., Keating, B., Kusaka, A., Lashner, J., Lee, A.~T., Limon, M., Mauskopf, P.~D., McMahon, J., Nati, F., Niemack, M.~D., Orlowski-Scherer, J.~L., Parshley, S.~C., Puglisi, G., Reichardt, C.~L., Salatino, M., Staggs, S., Suzuki, A., Vavagiakis, E.~M., Wollack, E.~J., Xu, Z., and Zhu, N., ``{Systematic uncertainties in the Simons Observatory: optical effects and sensitivity considerations},'' in [{\em Millimeter, Submillimeter, and Far-Infrared Detectors and Instrumentation for Astronomy IX}{\nolinebreak\hspace{0.1em}]},  Zmuidzinas, J. and Gao, J.-R., eds.,  {\bf 10708},  107083Y, International Society for Optics and Photonics, SPIE (2018).

\bibitem{AHuber2022}
Huber, A.~I., Chapman, S.~C., Sinclair, A.~K., Spencer, L.~D., Austermann, J.~E., Choi, S.~K., Devina, J., Gallardo, P.~A., Henke, D., Huber, Z.~B., Keller, B., Li, Y., Lin, L.~T., Niemack, M., Rossi, K.~M., Vavagiakis, E.~M., and Wheeler, J.~D., ``{CCAT-prime: optical and cryogenic design of the 850 GHz module for Prime-Cam},'' in [{\em Millimeter, Submillimeter, and Far-Infrared Detectors and Instrumentation for Astronomy XI}{\nolinebreak\hspace{0.1em}]},  Zmuidzinas, J. and Gao, J.-R., eds.,  {\bf 12190},  121901D, International Society for Optics and Photonics, SPIE (2022).

\bibitem{Datta_ARcoatings_2013}
Datta, R., Munson, C.~D., Niemack, M.~D., McMahon, J.~J., Britton, J., Wollack, E.~J., Beall, J., Devlin, M.~J., Fowler, J., Gallardo, P., Hubmayr, J., Irwin, K., Newburgh, L., Nibarger, J.~P., Page, L., Quijada, M.~A., Schmitt, B.~L., Staggs, S.~T., Thornton, R., and Zhang, L., ``Large-aperture wide-bandwidth antireflection-coated silicon lenses for millimeter wavelengths,'' {\em Appl. Opt.}~{\bf 52},  8747--8758 (Dec 2013).

\bibitem{huber_EoRSpec_optics_2022}
Huber, Z.~B., Choi, S.~K., Duell, C.~J., Freundt, R.~G., Gallardo, P.~A., Keller, B., Li, Y., Lin, L.~T., Niemack, M.~D., Nikola, T., Reichers, D.~A., Stacey, G., Vavagiakis, E.~M., and Zou, B., ``{CCAT-prime: the optical design for the Epoch of reionization spectrometer},'' in [{\em Millimeter, Submillimeter, and Far-Infrared Detectors and Instrumentation for Astronomy XI}{\nolinebreak\hspace{0.1em}]},  Zmuidzinas, J. and Gao, J.-R., eds.,  {\bf 12190},  121902H, International Society for Optics and Photonics, SPIE (2022).

\bibitem{Freundt_EoRSpec_2024}
Freundt, R.~G. and {CCAT Collaboration}, ``{CCAT: A status update on the EoR-Spec instrument module for Prime-Cam},'' in [{\em Millimeter, Submillimeter, and Far-Infrared Detectors and Instrumentation for Astronomy XII}{\nolinebreak\hspace{0.1em}]},  {\em Proc. SPIE} {\bf Paper number 13102-65} (2024).

\bibitem{AHuber_850GHz_2024}
Huber, A.~I. and {CCAT Collaboration}, ``{CCAT: Design and performance of densely packed, high-frequency, dual-polarization kinetic inductance detectors for the Prime-Cam 850 GHz Module},'' in [{\em Millimeter, Submillimeter, and Far-Infrared Detectors and Instrumentation for Astronomy XII}{\nolinebreak\hspace{0.1em}]},  {\em Proc. SPIE} {\bf Paper number 13102-2} (2024).

\end{thebibliography}
\bibliographystyle{spiebib} % makes bibtex use spiebib.bst

\end{document}